\documentclass[aps,prl,showpacs,showkeys,twocolumn]{revtex4-1}

\usepackage{mathptmx}
\usepackage{graphicx}

\newcommand{\bbm}[1]{\mathbf{#1}}

\begin{document}

\title{The triggering role of carrier mobility in the  fractional quantum Hall effect---evidence  in graphene\\}

\author{J. Jacak, L. Jacak}
\affiliation{Institute of Physics, Wroc{\l}aw University of Technology, Wyb. Wyspia{\'n}skiego 27, 50-370 Wroc{\l}aw, Poland}

\begin{abstract}
Recent experiments on suspended graphene layers have indicated the crucial role of carrier mobility in the competition between Laughlin collective state and insulating state, probably of Wigner-crystal-type. Moreover, the fractional quantum Hall effect (FQHE) in graphene has been observed at a low carrier density where the interaction is reduced as a result of particles dilution. This suggests that the interaction may not be as important  in the triggering of FQHE as expected based on the standard  formulation  of the composite fermions model. Here, the topological arguments are presented to explain these novel features of the FQHE in graphene and,  the  triggering role of carriers mobility in particular.

\end{abstract}

\pacs{05.30.Pr, 73.43.-f}
\keywords{graphene, FQHE, braid groups, composite fermions}

\maketitle

\subsection{Introduction}
 
While the integer quantum Hall effect (IQHE) is regarded as a single-particle and by topology conditioned phenomenon related to Landau level (LL) quantization (as has been especially convincingly argued in the topological quantum pump model of Laughlin \cite{laugh55}) with vital role played by impurity states \cite{klitzing1980}, the FQHE is essentially a collective state arrangement with primary role of the interaction. This is noticeable due to form of Laughlin wave function \cite{wlasne1} which turns out \cite{haldane,prange} to be an exact ground state for  2D system of charged particles at the presence of magnetic field  at lowest LL filling $1/p$ ($p$--odd) when the short range part of Coulomb interaction is taken  into account only (expressed by so-called Haldane pseudopotentials, i.e., matrix elements of Coulomb interaction in the base of relative angular momentum $m$ of electron pair; short range part of the interaction is limited by the $m=p-2$ Haldane term for the $p$-th FQHE state and the long range tail with $m>p-2$ does not influence significantly this state \cite{prange}). Despite the   
strong correlation effects the essential physics of the FQHE was successfully grasped upon an effectively single particle model of composite fermions (CFs) \cite{jain}, of only residually interacting particles associated with auxiliary field flux tubes each ($p-1$ flux quanta of the auxiliary field is assumed to be fixed in some way to each particle). These composite particles gain the statistics required by the Laughlin function as a result of the Aharonov-Bohm effect, when particles interchange together with the flux tubes fixed to them. The CF concept is commonly accepted because of its appealing single particle picture and  further modeling by variants of Chern-Simons field suitable to efficient calculations, supported with a good agreement with exact diagonalizations, especially inside the lowest LL (LLL) \cite{hon,jain2007}. 

Recent experimental investigations of the FQHE in graphene \cite{fqhe2,fqhe1} have shed, however,  new light on this correlated state and seem to go beyond explanative ability of CF treatment that concentrates solely on the interaction.  If one imagines CFs to be analogous to solid-state Landau quasiparticles dressed with the interaction, i.e., presuming that the  local flux tubes are a result of the interaction \cite{hon,jain2007}, one would lose  important topological effects and  encounter problems with new observations indicating that the carrier mobility (and not the interaction) plays a triggering role for the FQHE  in suspended graphene samples.

In the present letter we  revisit the  foundations of the composite fermion model and supply a topological explanation of the  FQHE that is in agreement with  recent observations in suspended graphene, thereby highlighting the crucial role mobility besides the interaction in FQHE.  Simultaneously topological conditioning of the FQHE Laughlin correlations is clarified.

\subsection{IQHE and FQHE in graphene---description of experimental results}

\begin{figure}[ht]
\centering
\scalebox{0.9}{\includegraphics{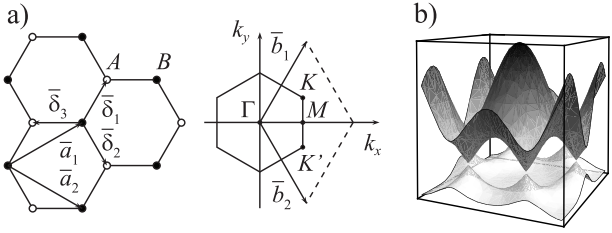}}
\caption{\label{grafenrys1} a)  A two-sublattices ($A$, $B$) triangular structure of graphene, $\vec{a}_1$, $\vec{a}_2$ are Bravais lattice vectors, $\vec{b}_1$, $\vec{b}_2$ are vectors of reciprocal lattice, b) the band structure of graphene, band type $\pi$, in a model of strong coupling, in compliance with the relation, $E_{\pm}({\bbm k}) = \pm t \sqrt{3+f({\bbm k})} - t'f({\bbm k})$, where $f({\bbm k})=2\cos(\sqrt{3}k_{y}a)+ 4\cos(\frac{\sqrt{3}}{2}k_{y}a)\cos(\frac{3}{2}k_{x}a)$, $t=2.7$ eV---hopping energy to the nearest neighbors (between sublattices, vectors $\vec{\delta}_i$), $t'=0.2t$---hopping energy to next-nearest neighbors (inside the sublattices), $a=1.42$ \AA, the ideal Dirac points $K$ and $K'$ for $t'=0$}
\end{figure}

A single-atom-thick layer of graphite (an allotropic form of carbon) known as graphene creates a hexagonal 2D structure, with a Bravais lattice with two vectors:
$$
\vec{a}_1=a(3,\sqrt{3})/2,\;\;\vec{a}_2=a(3,-\sqrt{3})/2,
$$
($a\simeq 0.142$ nm, distance between carbon atoms) with two carbon atoms per unit cell, Fig.\ref{grafenrys1} a.
This results in the creation of a double triangular lattice---a hexagonal structure of carbon atoms arranged in a honeycomb pattern, an outspread nanotube. Hybridized bonds $sp^2$ of the $\sigma$ type lead to a strong (with covalent bonds) two-dimensional structure, whereas $p$ orbitals perpendicular to the plane hybridize to type $\pi$ of the band structure (well described in the approximation of a strong coupling and with the inclusion of nearest neighbors and subsequent neighbors) with hole valley (valence band) and electron valley (conduction band) met in points $K$ and $K'$ at the border of a hexagonal Brillouin zone \cite{gr1,gr2},
Fig.\ref{grafenrys1} b.

\begin{figure}[ht]
\centering
\scalebox{1.0}{\includegraphics{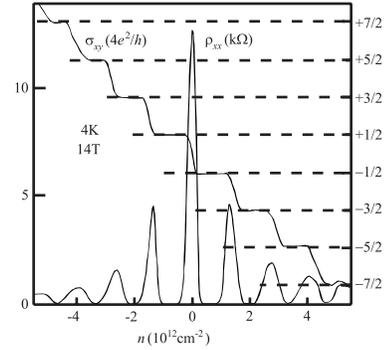}}
\caption{\label{grafenrys2} The IQHE in graphene in a concentration function (controlled with lateral gate voltage): the peak for $n=0$ indicates the existence of a Landau level at the Dirac point; electron and hole bands lead to symmetric IQHE  oscillation, plateaus $\sigma_{xy}$ correspond to half multiplicities of $4e^2/h$, according to the structure of the double-valley pseudospin {\scriptsize \textit{Source: A. H. Castro Neto, F. Guinea, N. M. R. Peres, K. S. Novoselov, and A. K. Geim, "The electronic properties of graphene," Rev. Mod. Phys. 81(1), p. 109, 2009}}}
\end{figure}

 Both bands meet in these points (non-gap semiconductor) and have a conical shape (near points $K,\;K'$, neglecting next-neighbors hopping), which means that the relation between energy and momentum (distance from points of contact) is linear with respect of momentum length. The appropriate band Hamiltionian (within the strong coupling approximation including the nearest neighbors and accounting for  both sublattices, numerated with an artificially introduced pseudospin) is formally equivalent to the description of relativistic fermions with zero rest mass ($E=\pm\sqrt{m_0^2v_F^4+p^2v_F^2}$, with $m_0=0$), described by Dirac equation with the velocity of light replaced by the Fermi velocity, $v_F\simeq c/300$ \cite{gr2,yang}. Therefore, the dynamics equation looks as follows,
$$ 
-i v_F \vec{\sigma} \cdot \nabla \Psi({\bbm r})=E\Psi({\bbm r}),
$$
where the Pauli matrix vector corresponds to the pseudospin structure related to two sublattices \cite{gr2,gr3} (wave functions are spinors in this structure). The zero mass of the Dirac fermions leads to numerous consequences and electron anomalies in the properties of graphene \cite{gr2,gr3,gr4,gr5}. For Dirac particles with zero rest mass, momentum uncertainty also leads to energy uncertainty (contrary to non-relativistic case, where the relation between uncertainty of position and momentum are independent from the relation of uncertainty of energy and time), which leads to the time evolution mixing together particle states with hole (anti-particle)  states for relativistic type dynamics. For zero-mass Dirac electrons the scaling of cyclotron energy is different as well ($\sim B^{1/2} $, and not $\sim B$, as in the case of non-relativistic particles). The value of this energy is also different, and larger by far (two orders of magnitude larger than the one corresponding in classical materials, i.e., it is [due to zero mass in Dirac point] as much as about 1000 K, for 10 T field), which allows to observe the IQHE in graphene even at room temperatures \cite{gr4,gr5}. There is, however, an anomalous IQHE observed here (for $\nu=\pm 4(n+1/2)$, or for $\pm 2, \pm 6, \pm 10, \ldots$ and at zero Landau level in the Dirac point, i.e., for zero energy; $\pm$ corresponds to particles and holes, respectively,  4 results from pseudospin/valley degeneration, 1/2 is associated with Berry's phase for pseudospin) \cite{gr4,gr5,yang}); cf. Fig. \ref{grafenrys2}, which is well-explained by the band structure leading to an effective Dirac description \cite{gr2,gr3,gr4,gr5,clure}. The  Klein paradox, referring to ideal tunneling of Dirac particles by rectangular potential barriers leads to extensive mobility of charge carriers in graphene, which is experimentally observed even near Dirac point (Fermi level at the border between electrons and holes). In this point, the density of charges is zero (and the zero Landau level is located here, employing both bands) \cite{gr2,gr4,gr5,yang}.

\begin{figure}[ht]
\centering
\scalebox{0.9}{\includegraphics{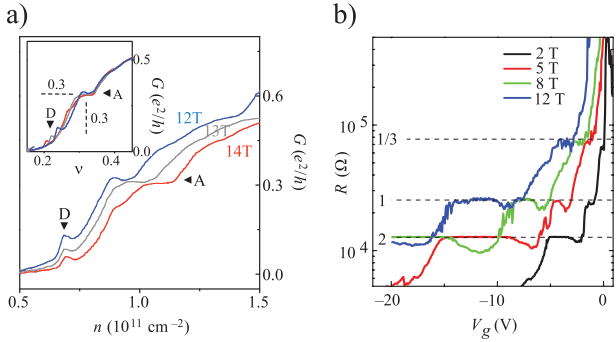}}
\caption{\label{grafenrys3} a) FQHE observation in suspended graphene for the filling 0.3 ($1/3$) in a field of $12-14$ T with the concentration of $10^{11}$cm$^{-2}$ and the mobility of 250000 cm$^2$V$^{-1}$s$^{-1}$, b) FQHE singularities in suspended graphene for the filling $\frac{1}{3}$ in a field of $2-12$ T with the concentration of $10^{10}$cm$^{-2}$ and the mobility of 200000 cm$^2$V$^{-1}$s$^{-1}$ {\scriptsize \textit{Source: a) X. Du, I. Skachko, F. Duerr, A. Liucan, and E. Y. Andrei, "Fractional quantum Hall effect and insulating phase of Dirac electrons in graphene," Nature 462, p. 192, 2009 b) K. I. Bolotin, F. Ghahari, M. D. Shulman, H. L. St{\"o}rmer, and P. Kim, "Observation of the fractional quantum Hall effect in graphene," Nature 462, p. 196, 2009}}}
\end{figure}

The search for states related to the FQHE in the case of Hall graphene measurements is particularly interesting. Despite using very strong magnetic fields (up to 45 T), FQHE was not detected in graphene samples deposited on a substrate of $SiO_2$ \cite{dur1}. In \cite{dur1} it was  noted, however, the emergence of additional plateaus  of IQHE for the fillings $\nu = 0,\pm 1, \pm 4$, indicating the elimination of spin-pseudospin degeneration (related to sublattices), as a result of increasing mass of Dirac fermions \cite{dur1}. Only after mastering the novel technique of the so-called suspended ultrasmall graphene scrapings with extreme purity and high mobility of carriers (beyond 200000 cm$^2$V$^{-1}$s$^{-1}$; note that high mobility is necessary also to observe the  FQHE  in the case of semiconductor 2D hetero-structures), was it possible to observe the FQHE in graphene at fillings $\nu=1/3$ and $-1/3$ (the latter for holes, with opposite polarization of  the gate voltage, which determines position Fermi level, either in the conduction band, or in the valence band) \cite{fqhe1,fqhe2}. Both papers report the observation of the FQHE in graphene for strong magnetic fields. In the paper \cite{fqhe1}, in a field of 14 T, for electron concentration of $10^{11}$/cm$^2$ and in the paper \cite{fqhe2}, in a field of 2 T, but for a concentration level smaller by one order  of magnitude---Fig.\ref{grafenrys3}. 

The FQHE in suspended graphene has been  observed at the temperatures around 10 K \cite{fqhe3}, and even higher (up to 20 K) \cite{fqhe6}. Authors  have argued that the critical  temperature elevation is related to the stronger electric interaction caused by the  lack of a dielectric substrate (with a relatively high dielectric constant in case of semiconductors, $\sim 10$) in the case of suspended samples. However, aspects that are likely  more important are high mobility value (with suppressed acoustic phonon interaction in ideal 2D system, in comparison to 3D case) and,  on the other hand, with very high cyclotron energy in graphene (i.e., the large energy gap between the incompressible states). The fractional Hall effect in graphene is also considered in relation to pseudospin structure in terms of symmetry SU(4) and SU(2) \cite{fqhe4}.

\begin{figure}[ht]
\centering
\scalebox{0.9}{\includegraphics{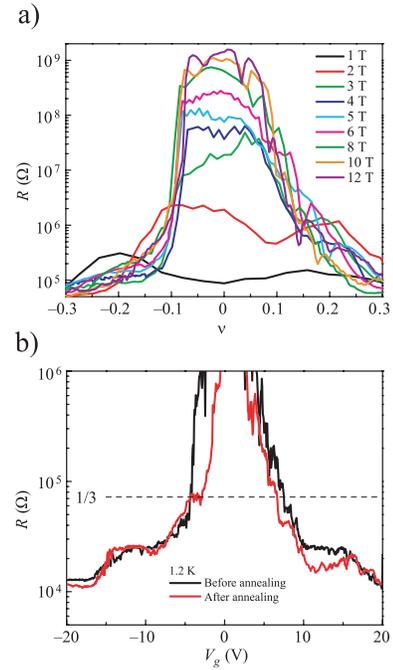}}
\caption{\label{grafenrys4}a) The emergence of an insulator state accompanying the increase in the strength of a magnetic field around the Dirac point, b) competition between the FQHE and the insulator state for the filling $-1/3$: annealing removes pollution which enhances the mobility and provides conditions for the emergence of a {\it plateau} for the FQHE {\scriptsize \textit{Source: K. I. Bolotin, F. Ghahari, M. D. Shulman, H. L. St{\"o}rmer, and P. Kim, "Observation of the fractional quantum Hall effect in graphene," Nature 462, p. 196, 2009}}}
\end{figure}

In the papers \cite{gr5,clure} the competition between the FQHE state with the insulator state near the Dirac point has also been demonstrated, corresponding to a rapidly decreasing carrier concentration (and thus reducing interaction role at larger separation of carriers)---Fig.\ref{grafenrys4}.
 The most intriguing observation is that one \cite{fqhe2} demonstrating an influence of  annealing---in Fig.\ref{grafenrys4} b it is shown that FQHE occurs in the same sample originally insulating, upon same conditions, but after the annealing process enhancing mobility of carriers due to impurities reduction. This effect directly demonstrates the triggering role of carriers mobility in FQHE state arrangement. 

\subsection{Quasiclassical quantization of the magnetic field flux for composite fermions}

The relevant topological theory of 2D charged multiparticle systems under strong magnetic field was  developed earlier in \cite{jac1,jac2}. The related main points are as follows: 
\begin{itemize}
    \item{for fractional fillings of the LLL classical cyclotron orbits are too short for particle exchanges (as in the 2D case, long range spiral  motion is impossible)}
\item{the exchanges are necessary to create a collective state such as the FQHE---thus the cyclotron radius must be enhanced (in the model of Jain's CFs \cite{jain} this enhancement is achieved by the artificial addition of flux tubes  directed oppositely to the external field; in the model of Read's vortices \cite{vor3}, reduction of the cyclotron radius is attained by the depletion of the local charge density---both this tricks are, however, effective and not supported in topological terms);}
\item{the topologically justified way to enhance cyclotron radius is with the use of multilooped cyclotron trajectories related to multilooped braids that describe elementary particle exchanges in terms of braid groups (the resulting cyclotron braid subgroup is generated by $\sigma_i^p$, $i=1,...,N$, $\sigma_i$ are generators of the full braid group \cite{jac1,jac2})---in 2D all loops of multilooped trajectory must share together the same total external magnetic field flux, in contrast to  the 3D case, and this is a reason of enhancement of all loops dimensions effectively fitting exactly to particle separation at LLL fillings $1/p$, ($p$-odd); } 
\item{in accordance with the rules of path integration for non-simply-connected configuration spaces \cite{lwitt},  one dimensional unitary representations (1DURs) of a corresponding braid group define the statistics of the system---in the case of multilooped braids, naturally assembled into the cyclotron subgroup \cite{jac2,EPL}, one arrives in this way at the statistical properties required by the Laughlin correlations (these 1DURs are $\sigma_i^p\rightarrow e^{ip\alpha},\;\alpha \in [0,2\pi)$, CFs correspond to $\alpha=\pi$);}
\item{the interaction is important for properly determining the cyclotron braid structure because its short range part  prevents the particles from approaching one another closer than the distance given by the density.}\end{itemize}

This completes a topological derivation of  FQHE statistics  without need to model it by  fictitious auxiliary elements like flux tubes, and proves that CFs are not quasiparticles dressed with interaction or complexes with local fluxes, but are rightful quantum 2D particles assigned with Laughlin statistics determined  by 1DURs of the appropriate cyclotron braid subgroup.

Even though the auxiliary field  flux-tubes fixed to CFs  do not exist and only mimic the cyclotron braid structure, they are very useful because of the simplicity of the model and of exploitation of the Chern-Simons field to efficient calculation scheme. Numerical results have
indicated the high accuracy of this model, at least for LLL fractional states, but also its convenience in higher LLs.  Since the model of CFs with attached rigid flux quanta  works so well (as evidenced by the exact diagonalizations), the multilopped classical braid structure must be repeated by quasiclassical wave-packet trajectories (and then with quantized fluxes, which was not, however, a rule for classical  trajectories). Note that the flux quantization is the quasiclassical property as it needs a trajectory definition (the carries mobility also has the similar quasiclassical background). 

The wave packets corresponding to the  quasiclassical dynamics are  related to the collective character of a multiparticle system. For LLL of noninteracting system the group velocity of any packet is zero due to states degeneration. Interaction removes, however, this degeneracy and provides packet dynamics.  The collective movement minimizes kinetic energy, whereas the interaction favors localization (localization causes related increase in kinetic energy). Therefore, the collective dynamics seems to prefer the quasi-classical movement of packets along periodical closed trajectories (as a rule at the magnetic field presence \cite{eliu}), which then must, however, embrace  quantized external magnetic field fluxes. This suggests a role for collectivization in the energetic preference of wave packets traversing closed trajectories in correspondence with the classical cyclotron description, including the multilooped braid picture \cite{jac2}.

This description appears to  be in accordance with the FQHE observations in graphene (described above), which are found at a low carrier density, and therefore accompany their dilution and the resulting reduction of interaction. Thus, the interaction is not the main factor initiating the FQHE, as was previously expected in view of the standard model of composite fermions, if one treated the dressing of fermions with localized flux tubes as a result of just the interaction itself \cite{hon,jain2007}.
Carrier mobility refers to semiclassical wave packet dynamics in terms of the drift velocity in an electric field and the classical Hall effect and reflects various channels of scattering phenomena beyond the simple model that includes only Coulomb interaction to free particles in the magnetic field. But topology arguments in the 2D case strongly prefer high mobility which is  required for real wave packets   to traverse multilooped trajectories. Semiclassical wave packets even at presence of interaction and scattering,  manifest periodic dynamics \cite{eliu} and in the case of the multilooped trajectory structure with enhanced radii,  higher mobility is required, as has been experimentally demonstrated.

\subsection{Cyclotron braid group approach to FQHE in graphene}

From the  cyclotron group pont of view, experimental results on FQHE in graphene \cite{fqhe1,fqhe2,fqhe3,fqhe6,fqhe4} seem to be compliant with the expectations of the braid description. In the case of graphene, the specific band structure (one of a gapless semiconductor) with conical Dirac bands leads to simultaneous participation (in the case of Dirac point) of both bands---of holes and of electrons, which combined with the massless character of Dirac fermions manifests itself through an anomalous IQHE (further corrected  in very strong fields as a result of eliminating spin and two-valley degeneration) \cite{gr4,gr5,dur1}. Controlling lateral gate voltage (within the range up to 10 V \cite{fqhe1}) allows regulating the density of carriers at a constant magnetic field. One should therefore expect  that at relatively  small densities of carriers (electrons, or symmetrical holes at reverse voltage polarization), the  cyclotron orbits will be too short to prevent braid exchanges of particles at a sufficiently strong magnetic field---although weaker for smaller concentrations---and experimental observations have supported  exactly  this prediction \cite{fqhe1,fqhe2}. For low concentration, while closing on the Dirac point, one may expect that too strong fields would exceed the stability threshold of the FQHE state in competition with the Wigner crystal (accounting for a specific character of this competition in the case of massless Dirac fermions in comparison to traditional semiconductor 2D structures, cf. Ref. \cite{wig-no}) and that corresponds to the emergence of the insulating state near the Dirac point in a strong magnetic field \cite{fqhe5}. In the case of the hexagonal structure of graphene, electron (or hole) Wigner crystallization may exhibit interference between the triangular crystal sublattices, and inclusion of the resonance (hopping) between these two sublattices may cause blurring of the sharp transition to the insulator state, which seems compliant with observations (Fig. \ref{grafenrys4}).

\subsection{Comments}

Although   the real dynamics of quasiclassical wave packets is beyond the description ability  in the framework of a simplified 2D multiparticle charge system upon magnetic field, some general qualitative conclusions regarding the topological character can be drawn. In the case of the fractional filling of the LLL, when classical cyclotron orbits are too short for braid exchanges, the multilooped trajectories occur. Via linkage of this classical cyclotron dynamics with quasiclassical wave packet trajectories, supported by a success of CF model with rigid flux quanta attached to particles, one can expect also multilooped structure of real quasiclassical wave packets dynamics. With larger radii  these multilooped orbits strongly favor higher mobility of carriers and this has been confirmed experimentally in suspended graphene. This suggests that the mobility is of primer significance for FQHE formation in competition with localized electron states like insulating Wigner-crystal-type-state. This picture seems to agree also with already predicted \cite{wig10,wig119} destabilization of Laughlin state in semiconductor heterostructures at $p>9$ in favor to Wigner crystal, when visibly too many loops begin to be energetically unfavorable. Experiments with graphene have indicated  that not only interaction influences stability of FQHE state, which agrees with the topological explanation of CFs properties, thereby demonstrating that the previous concept of flux tubes generated by the interaction itself is insufficient.          

%

\end{document}